\documentclass[aps,pre,reprint,showpacs,groupedaddress]{revtex4-1}
 \usepackage{amsmath}
 \usepackage{graphicx}
 \usepackage{bmpsize}
\begin{document}

\title{Mean field approach for diffusion of interacting particles}

\author{G. Su\'arez}
\email{gsuarez@mdp.edu.ar}
\author{M. Hoyuelos}
\email{hoyuelos@mdp.edu.ar}
\author{H. M\'artin}

\affiliation{Instituto de Investigaciones F\'isicas de Mar del Plata (IFIMAR -- CONICET)}
\affiliation{Departamento de F\'isica, Facultad de Ciencias Exactas y Naturales,
             Universidad Nacional de Mar del Plata.\\
             De\'an Funes 3350, 7600 Mar del Plata, Argentina}

\date{\today}

\begin{abstract}
A nonlinear Fokker-Planck equation is obtained in the continuous limit of a one-dimensional lattice with an energy landscape of wells and barriers. Interaction is possible among particles in the same energy well. A parameter $\gamma$, related to the barrier's heights, is introduced. Its value is determinant for the functional dependence of the mobility and diffusion coefficient on particle concentration, but has no influence on the equilibrium solution. A relation between the mean field potential and the microscopic interaction energy is derived. The results are illustrated with classical particles with interactions that reproduce fermion and boson statistics.
\end{abstract}


\maketitle

\section{Introduction}
The complete description of a large number of interacting particles (many-body system) is a problem that usually exceeds numerical or analytical capabilities. Mean field theory allows the description of such a system by analyzing the behaviour of only one particle subjeted to an average force produced by the rest of the particles. The mean field force depends on statistical properties of the rest of the particles that, for self consistency, are equal to the properties of the particle considered before \cite[p. 131]{chandler}. This is the origin of the nonlinear character of the resulting description \cite[p. 3]{frank}. The nonlinearities reflect the interactions between particles. For example, Kaniadakis and Quarati \cite{kania} introduced a nonlinear Fokker-Planck equation for the particle density in a classical system that reproduce quantum statistics. An appropriate choice of the transition probabilities, that depend on the particle density, gives rise to Fermi-Dirac or Bose-Einstein distributions in equilibrium.
Quantum effects can be reproduced in a classical system with an interaction potential, also called statistical potential \cite[p. 124]{pathria}. For example, the Pauli exclusion principle for fermions is analogous to a potential that becomes infinite when two particles occupy the same state. This hard-core interaction is the cause of a nonlinear term in the corresponding diffusion equation, see \cite{kania}, \cite[p. 280]{frank} or Eq. (10) in \cite{suarez}. A general approach for the derivation of a nonlinear Fokker-Planck equation starting form the Master equation can be found in \cite{curado,nobre}. More examples can be found in \cite{frank}.

Many researchers have devoted attention to the problem of diffusion with interaction, in many cases motivated by the seminal work of Batchelor \cite{batchelor}. The following list is a limited and partial sample of related references: \cite{murphy,felderhof,finsy,ohtsuki,lekkerkerker,cichocki,
guidoni,aranovich,savelev,chen,nelissen,simpson,bruna,beker,arita}. 

The problem that we wish to address, in a quite general perspective, is collective diffusion of interacting particles. 
We use a one dimensional lattice, but the results can be easily extended to higher dimensions. In our analysis, only the collective or transport diffusion coefficient is involved (we do not analyze here single particle diffusion, or self-diffusion of tagged particles).

The restriction on the interaction is that it is local: it takes effect only among particles in the same site. In other words, the interaction range is smaller than the lattice spacing. We apply a mean field approximation assuming that the evolution of the system can be obtained analyzing the behavior of only one particle. We consider that this particle is subjected to a mean field potential $V_i$ that depends on the number of particles in the site, $n_i$, where $i$ is the lattice site index. The variation of the particle number is smooth so that the continuous limit can be applied, and a nonlinear Fokker-Planck equation is obtained. The equilibrium solution is completely determined by the mean field interaction potential and, if present, an external potential. This is not the case for the non-equilibrium behavior. As we show in the next sections, it depends on an additional parameter, $\gamma$, that determines if the transition probability between neighboring sites depends on the potential in the source site, in the target site, or on a mixture of both potentials.

\section{Transition probabilities}

In the one-particle picture, the energy associated to a particle in site $i$ is given by
\begin{equation}
E_i = V_i + U_i, \label{energy} 
\end{equation}
where $V_i$, the internal or mean field potential, is an abbreviation of $V(n_i)$, and $U_i$ is an external potential. We can consider $V_i$ as the function of $n_i$ that, in equilibrium, satisfies the relation $n_{\textrm{eq},i} \propto \exp\{-\beta [V(n_{\textrm{eq},i}) +U_i]\}$, i.e., it is a one-particle effective potential that satisfies Boltzmann statistics in equilibrium. 

The interaction energy of a configuration of $n_i$ particles is given by a function $\Phi(n_i)$. The relevant problem of determining the relation between the mean field potential $V_i$ and the interaction energy $\Phi(n_i)$ is addressed in Sect.\ \ref{meanfield}.  The interaction is local in the sense that particles interact only when they are in the same site. 
 
In the one-particle picture, the detailed balance condition gives a relation between transition probabilities:
\begin{equation}
e^{-\beta E_{i+1}} W_{i+1,i} = e^{-\beta E_i} W_{i,i+1},
\label{detbal}
\end{equation}
where $W_{i,i+1}$ is the transition probability from site $i$ to site $i+1$. Using the words of Derrida \cite{derrida}, it is a ``straightforward generalization'' to consider that \eqref{detbal} also holds for the transition probabilities out of equilibrium (it is clear that this does not mean that detailed balance holds out of equilibrium, since in general $n_i \neq \textrm{const} \times e^{-\beta E_i}$).

Detailed balance is not enough to determine the transition probabilities. In order to obtain an expression for them, we assume that they can be written as a combination of exponentials of $V_i$, $V_{i+1}$, $U_i$ and $U_{i+1}$. In the paragraphs below we present a physical interpretation of the result that also serves as a justification of this assumption. We have
\begin{equation}
W_{i,i+1} = P\,\exp\left[-\beta\, (\gamma\, V_{i+1} + \gamma' \, V_i + \alpha \,U_{i+1} + \alpha'\,U_i)\right],
\label{trans0}
\end{equation}
where $\gamma$, $\gamma'$, $\alpha$, $\alpha'$ and $P$ may depend on position $i$; the subindex is omitted in order to lighten the notation. We assume that the system has an inversion symmetry. It implies that the reversed transition probability is obtained from \eqref{trans0} by exchanging $i \leftrightarrow i+1$:
\begin{equation}
W_{i+1,i} = P\,\exp\left[-\beta\, (\gamma\, V_i + \gamma' \, V_{i+1} + \alpha \,U_i + \alpha'\,U_{i+1})\right],
\label{trans1}
\end{equation}
It can be shown that the combination that fulfills detailed balance is $\gamma'=\gamma-1$ and $\alpha'=\alpha-1$. Then,
\begin{equation}
W_{i,i+1} = P\, e^{- \beta \left[(\gamma-1)V_i + \gamma V_{i+1} + (\alpha-1) U_i +\alpha U_{i+1} \right]}.
\label{transprob}
\end{equation}

\begin{figure}
\includegraphics[width = 0.45\textwidth]{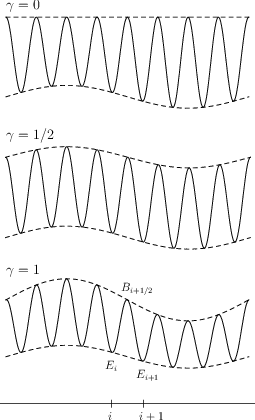}
\caption{Energy landscape of wells ($E_i$) an barriers ($B_{i+1/2}$). From to to bottom: $\gamma=0$, $1/2$ and 1 $\forall i$. In this illustration, $C$ is constant and $U_i=0 \; \forall i$.}
\label{potential}
\end{figure}

A physical interpretation of $\gamma$ and $\alpha$ parameters is related to the barriers height between neighboring sites. Le us consider that particles can occupy discrete sites in a lattice with a potential that has a continuous shape, as shown in Fig. \ref{potential}. Between sites $i$ and $i+1$, the potential has a maximum value of $B_{i+1/2}$. A particle that jumps from $i$ to $i+1$ has to overcome a barrier $B_{i+1/2} - E_i$, and the transition probability is given by
\begin{equation}
W_{i,i+1} = \nu e^{-\beta(B_{i+1/2} - E_i)},
\label{barrier}
\end{equation}
where $\nu$ is the number of jump attempts per unit time. The potential maxima depend, on one hand, on the characteristics of the substratum, that contributes with a term $C$. They also may depend on the values of the interaction potential at both sides of the barrier. This dependence is symmetrical (the same for both sides) and its influence is represented by a term $\gamma (V_i + V_{i+1})$, the choice of the name of $\gamma$ parameter advances that it is the same than the one introduced in \eqref{transprob}. Finally, the external potential also has an influence on the barriers. It is given by an additional parameter $\eta \in (0,1)$: $\eta U_i + (1-\eta) U_{i+1}$. The influence of the external potential on the barriers is always present. Parameter $\eta$ is an interpolation factor between sites $i$ and $i+1$ for this influence.  Taking these arguments together, the potential maximum is
\begin{equation}
B_{i+1/2} = C +\gamma(V_i + V_{i+1}) + \eta U_i + (1-\eta) U_{i+1}.
\end{equation}
Replacing in \eqref{barrier}, and considering that $P = \nu e^{-\beta C}$, we obtain 
\begin{equation}
W_{i,i+1} = P e^{-\beta\left[ (\gamma-1)V_i + \gamma V_{i+1} + (1-\eta) (U_{i+1} - U_i) \right]}.
\label{transprob2}
\end{equation}
Let us compare this with Eq. \eqref{transprob}. Since the values of $U_i$ and $U_{i+1}$ are, in principle, arbitrary, we have that $\alpha-1=-1+\eta$ and $\alpha = 1-\eta$ and, therefore, $\alpha=\eta = 1/2$. This result is actually a consequence of the inversion symmetry assumed in \eqref{trans1}. Finally, the expression for the transition probability is 
\begin{equation}
W_{i,i+1} = P e^{-\beta\left[ (\gamma-1)V_i + \gamma V_{i+1} + \Delta U/2 \right]},
\label{transprob3}
\end{equation}
with $\Delta U = U_{i+1} - U_i$.

The previous arguments allow us to construct an energy landscape that gives physical meaning to the parameters and variables involved in the transition probability. In this picture we are using the H\"anggi \cite{hanggi} interpretation of Lagevin equation with multiplicative noise, or barrier model \cite{sokolov}, for which the particle current,  without external potential and in a continuous space, is $J = -D(x) \frac{\partial n}{\partial x}$ (for other interpretations, the spatial dependent diffusion coefficient is inside the space derivative \cite{sokolov}).

In principle, $\gamma$ can take any real value. We conjecture that, from a physical point of view, the pertinent values of $\gamma$ are in the range $[0,1]$. Typical situations are shown in Fig. \ref{potential}. For $\gamma=0$, the transition probability depends on the origin potential:
\begin{equation}
W_{i,i+1} = P e^{-\beta(-V_i + \Delta U/2)}\quad\quad (\gamma=0).
\end{equation}
For $\gamma=1$, the transition depends on the target potential:
\begin{equation}
W_{i,i+1} = P e^{-\beta(V_{i+1} + \Delta U/2)}\quad\quad (\gamma=1).
\label{gamma1}
\end{equation}
And for $\gamma=1/2$ we have an intermediate case that is a frequent choice in Monte Carlo (MC) simulations of diffusion processes; the transition depends on the energy difference between target and origin potentials:
\begin{equation}
W_{i,i+1} = P e^{-\beta(\Delta V + \Delta U)/2}\quad\quad (\gamma=1/2),
\end{equation}
with $\Delta V = V_{i+1} - V_i$.

\section{Current and Fokker-Planck equation}

The current $J$ between sites $i$ and $i+1$ is
\begin{equation}
J = n_i\, W_{i,i+1} - n_{i+1} \, W_{i+1,i}.
\label{current}
\end{equation}
We replace \eqref{transprob3} in \eqref{current}. Now we turn to a continuous description in which $n_i$ is replaced by $n(x)$, $V_i$ by $V(n(x))$, and $U_i$ by $U(x)$, with $x=a i$; we approximate $(n_{i+1}-n_i)/a \simeq \frac{\partial n}{\partial x} $, $(V_{i+1}-V_i)/a \simeq \frac{d V}{d n} \frac{\partial n}{\partial x}$ and $(U_{i+1}-U_i)/a \simeq \frac{\partial U}{\partial x}$. We call $D_0 = Pa^2$ the free diffusion coefficient. After some algebra, we obtain
\begin{equation}
J\,a = -\mu\, \frac{\partial U}{\partial x}\, n -  D\, \frac{\partial n}{\partial x},
\label{ccurrent}
\end{equation}
where the mobility $\mu$ and the diffusion coefficient $D$ are
\begin{align}
\mu & = \beta D_0\,e^{-\beta(2\gamma-1)V} \label{mu}\\
D & = D_0\,e^{-\beta(2\gamma-1)V} \left( \beta n\frac{dV}{dn} + 1\right)
\label{diffcoef}
\end{align}
In the previous derivation the validity of the Ginzburg criterion for a mean field theory is assumed: the fluctuations are small enough so that $(\langle n_i^2 \rangle - \langle n_i\rangle^2)/\langle n_i^2 \rangle \ll 1$. 
The relation between mobility and diffusion coefficient
\begin{equation}
D = \mu \beta^{-1} \left( \beta n\frac{dV}{dn} + 1\right).
\end{equation}
does not depend on $\gamma$. In the absence of interaction, we recover the Einstein relation: $D = \mu \beta^{-1}$.

The resulting Fokker-Planck equation
\begin{equation}
\frac{\partial n}{\partial t} = \frac{\partial}{\partial x}\left( \mu\, \frac{\partial U}{\partial x}\, n +  D\, \frac{\partial n}{\partial x} \right)
\label{fp}
\end{equation}
is nonlinear because of the dependence of $\mu$ and $D$ on $n$ and $V(n)$. It is a free energy Fokker-Planck equation with Boltzmann statistics, see chapter 5 in \cite{frank}. If we consider the notation used in Eq.\ (5.4) of reference \cite{frank}, we find the following correspondence: $n \rightarrow P$, $\beta U \rightarrow U_0/Q$, $\beta V \rightarrow \frac{\delta U_\textrm{NL}}{\delta P}/Q$ and $D_0 \rightarrow Q$. The main novelty of our approach is the introduction of $\gamma$ parameter ---that plays a relevant role in the system's dynamics--- and its physical interpretation in the energy landscape.

Identifying the zero current state with equilibrium, it is easy to see that the equilibrium concentration is $n_\textrm{eq} \propto e^{-\beta(V + U)}$. The proportionality constant can be written as
\begin{equation}
n_\textrm{eq} = e^{-\beta(V + U - \mu_c)},
\label{neq}
\end{equation}
where we can identify $\mu_c$ with the chemical potential. As expected, $\gamma$ plays no role in the equilibrium solution. Its influence is present in the dynamics through the dependence of $\mu$ and $D$ on $\gamma$. Let us also note that the $\gamma$ parameter has influence only when an interaction potential is present.

\section{The mean field potential}
\label{meanfield}

Instead of speaking about the energy of one particle, as in \eqref{energy}, let us consider the energy of the configuration of $n_i$ particles:
\begin{equation}
\epsilon_i = \Phi(n_i) + n_i\, U_i,
\end{equation}
where $\Phi(n_i)$ is the interaction energy of the configuration of $n_i$ particles. The question that lies at the core of a mean field theory is: what is the relation between $V$ and $\Phi$?

In order to answer this question we have to appeal to the microscopic description given by the grand partition function for classical particles. Since there is not interaction between different lattice sites, it can be written as
\begin{equation}
\Xi = \prod_i Z_i
\end{equation}
with
\begin{equation}
Z_i = \sum_{n_i=0}^\infty \frac{1}{n_i!} e^{-\beta(\epsilon_i-n_i\mu_c)} = \sum_{n_i=0}^\infty \frac{1}{n_i!} e^{-\delta_i n_i - \beta \Phi(n_i)}
\label{partfunc}
\end{equation}
where, for simplicity, we introduced $\delta_i = \beta(U_i - \mu_c)$. 
We can obtain the equilibrium mean value $\langle n_i \rangle$ with
\begin{equation}
\langle n_i \rangle = -\frac{\partial \ln Z_i}{\partial \delta_i}
\end{equation}
and express this result as a function of $\delta_i$ (for a given value of $\beta$): $\langle n_i \rangle = f_\beta(\delta_i)$. This means that we can consider $\delta_i$ as a parameter that allows us to scan the possible values of $\langle n_i \rangle$. From an experimentalist point of view, we can apply a varying external potential and measure the possible values of $\langle n_i \rangle$. We will restrict ourself to situations in which $f_\beta$ is invertible, i.e., we can write $\delta_i = f^{-1}_\beta(\langle n_i \rangle)$, this precludes the possibility of phase transitions, for which, for a given $\delta_i$, there could be more than one value of $\langle n_i \rangle$.

The connection with the description of the previous section, based on the nonlinear Fokker-Planck equation, is given by the fact that $\langle n_i \rangle$ should be equal to $n_\textrm{eq}$ in $x=ai$. Then, using \eqref{neq}, we have
\begin{equation}
\langle n_i \rangle = e^{-\beta V - \delta_i}.
\label{meannv}
\end{equation}
In the same way that we considered that the transition probabilities satisfy the detailed balance relation even when the system is out of equilibrium, we can consider that the relation between the mean field potential $V$ and the particle concentration that we can derive from the previous equation also holds out of equilibrium. This can be justified using the local equilibrium assumption. For simplicity, we replace $\langle n_i \rangle$ by $n$.  Therefore, from \eqref{meannv} we obtain
\begin{equation}
V(n) = -\beta^{-1} \left( \ln n + f^{-1}_\beta(n)\right).
\label{vvsphi0}
\end{equation}

The most simple situation is zero interaction; in this case $n=e^{-\delta_i}$ and $f^{-1}_\beta(n) = -\ln n$. 
A more direct relation between $V$ and $\Phi$ can be derived from \eqref{meannv}; it also clarifies its physical meaning. We have
\begin{align}
e^{-\beta V} &= - e^{\delta_i} \frac{\partial \ln Z_i}{\partial \delta_i} 
= \frac{e^{\delta_i}}{Z_i} \sum_{n_i=0}^\infty \frac{n_i}{n_i!} e^{-\delta_i n_i - \beta \Phi(n_i)} \nonumber \\
&= \frac{1}{Z_i} \sum_{n_i=1}^\infty \frac{1}{(n_i-1)!} e^{-\delta_i (n_i-1) - \beta \Phi(n_i)} \nonumber \\
&= \frac{1}{Z_i} \sum_{n_i'=0}^\infty \frac{1}{n_i'!} e^{-\delta_i n_i' - \beta \Phi(n_i'+1)} \nonumber \\
&= \frac{1}{Z_i} \sum_{n_i'=0}^\infty e^{-\beta (\Phi(n_i'+1)-\Phi(n_i'))}\frac{1}{n_i'!} e^{-\delta_i n_i' - \beta \Phi(n_i')} \nonumber \\
&= \left\langle e^{-\beta (\Phi(n_i+1)-\Phi(n_i))} \right\rangle
\label{vvsphi}
\end{align}

We can interpret this result using the Jarzynski equality \cite{jarzynski}, that gives a relation between the Helmholtz free energy variation $\Delta F$ and the applied work $W$:
\begin{equation}
e^{-\beta \,\Delta F} = \left\langle e^{-\beta \,W} \right\rangle.
\label{jarzy}
\end{equation}
Comparing \eqref{vvsphi} and \eqref{jarzy}, we can see that the mean field potential $V$ at a given point is equal to the free energy change when the work needed to increase the number of particles by one, $W = \Phi(n_i+1)-\Phi(n_i)$, is applied to that point.

\section{Hard core interaction}

Hard core interaction is the prototypical case study. Phenomenological arguments to determine the transition probabilities are usually presented \cite{kania,simpson,arita,spohn}. Without the presence of an external field, the transition probability from $i$ to $i+1$ is proportional to the quantity $1-n_{i+1}$, that indicates if the target site is free to be occupied by the incoming particle:
\begin{equation}
W_{i,i+1} = P \, (1-n_{i+1}).
\label{transphen}
\end{equation}
Since the transition probability depends on the target site, we have $\gamma=1$. From Eq. \eqref{gamma1} we obtain $V_{i+1} = -\beta^{-1}\ln (1-n_{i+1})$. In the continuous limit, and neglecting fluctuations, 
\begin{equation}
V(n) = -\beta^{-1}\ln (1-n).
\label{vhc}
\end{equation}
The resulting equilibrium concentration, that is obtained replacing \eqref{vhc} in \eqref{neq}, corresponds to Fermi-Dirac statistics: $n_\textrm{eq} = 1/(e^{\beta (U-\mu_c)} + 1)$.

On the other hand, using the interaction energy
\begin{equation}
\Phi(n_i) = \left\{ \begin{array}{ll}
0 & \mbox{if } n_i= 0, 1 \\ 
\infty & \mbox{if } n_i \ge 2 
\end{array} \right.
\end{equation}
in \eqref{vvsphi}, we can arrive to the same result \eqref{vhc}. This procedure does not appeal to an a priori definition of the transition probabilities with the shape of \eqref{transphen}, and makes a clear distinction between the mean field potential $V$ and the microscopic interaction energy $\Phi$.

Several results obtained from diffusion in a lattice indicate that the hard core interaction does not have any effect on the collective diffusion coefficient \cite{kania,suarez,simpson,arita}:
\begin{equation}
D_{\textrm{HC}} = D_0. \label{diffconst}
\end{equation}
(Of course, it \textit{does} have an effect in the single particle diffusion coefficient \cite{richards,suarez2}.) This result is obtained from the expression for the diffusion coefficient \eqref{diffcoef} using \eqref{vhc} for the mean field potential and $\gamma=1$. 

Therefore, in our description based on the $\gamma$-dependent energy landscape, the result \eqref{diffconst} corresponds to a constant value of $\gamma$ equal to 1. But this is not the only possibility. For example, for $\gamma = 1/2$ and the same hard core interaction we obtain
\begin{equation}
D_\textrm{HC} = \frac{D_0}{1-n} \quad\quad (\gamma=1/2).
\end{equation}
It is interesting to note that a similar dependence of the diffusion coefficient on the concentration has been obtained for hard spheres in a continuous space (see, for example, \cite{batchelor,finsy,felderhof,bruna}). More complex expressions of the diffusion coefficient can be obtained if a $\gamma$ parameter that depends on the concentration is considered. In all cases the interaction, and the equilibrium solution, is the same.

Now we consider the situation in which multiple occupancy is allowed with a maximum number of particles $N$. Let us call $m_i$ the number of particles in site $i$ to distinguish this case from the one of the previous paragraphs. And let us consider that the average concentration is given by $\langle m_i \rangle = N \langle n_i \rangle$. In terms of the partition function:
\begin{equation}
\langle m_i \rangle = - N \frac{\partial \ln Z_i}{\partial \delta_i} = - \frac{\partial \ln Z_i^N}{\partial \delta_i}.
\end{equation}
Then, the partition function for the multiple occupancy case is
\begin{equation}
Z'_i = (1+e^{-\delta_i})^N = \sum_{m_i=0}^N e^{-\delta_i m_i} \left( \begin{array}{c}
N \\ 
m_i
\end{array} \right).
\end{equation}
Using the definition of the partition function, and comparing with the previous equation, we obtain the interaction energy for this case:
\begin{equation}
\Phi'(m_i) = \left\{ \begin{array}{ll}
-\beta^{-1} \ln \frac{N!}{(N-m_i)!} & \mbox{if } 0 \leq m_i \leq N \\ 
\infty & \mbox{if } m_i > N
\end{array} \right.
\end{equation}
The mean field potential is 
\begin{equation}
V(m) = -\beta^{-1} \ln(N-m).
\end{equation}
By increasing the value of $N$ we increase the mean number of particles in each site and reduce fluctuations, a procedure that favors the conditions for the validity of the Fokker-Planck equation. 

\begin{figure}
\includegraphics[width = 0.45\textwidth]{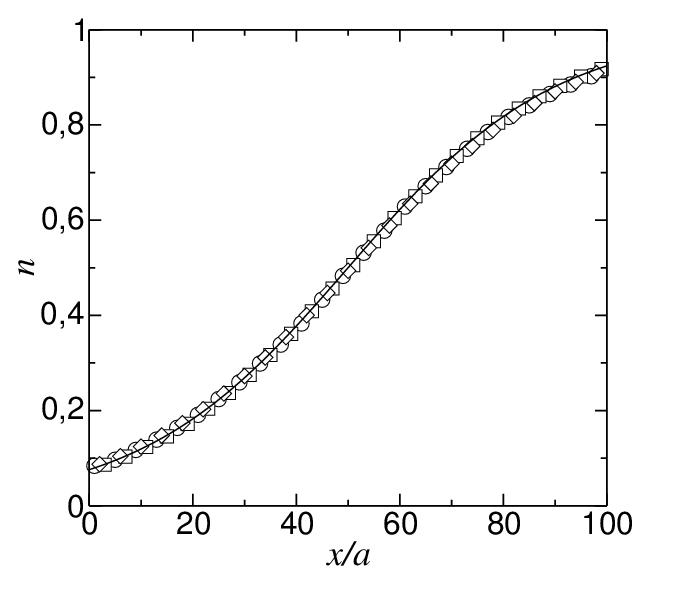}
\caption{Equilibrium concentration $n$ for hard core interaction with $U(x) = -F\,x$, $\beta F a=0.05$ and total concentration $0.5$. The curve corresponds to the Fermi-Dirac distribution and the dots to Monte Carlo simulations with different values of $\gamma$: $\gamma=0$ (circles), $\gamma=1/2$ (squares) and $\gamma=1$ (diamonds). The same distribution is attained for all values of $\gamma$. Simulations were made with multiple occupancy of sites: $n_i=m_i/N$. Number of samples: 1000; MC steps for each sample: $10^7$ ($\gamma = 0$ and 1, $N=100$) or $10^8$ ($\gamma=1/2$, $N=500$); fixed boundary conditions; system size $L=100 a$.}
\label{eqfemions}
\end{figure}

\begin{figure}
\includegraphics[width = 0.45\textwidth]{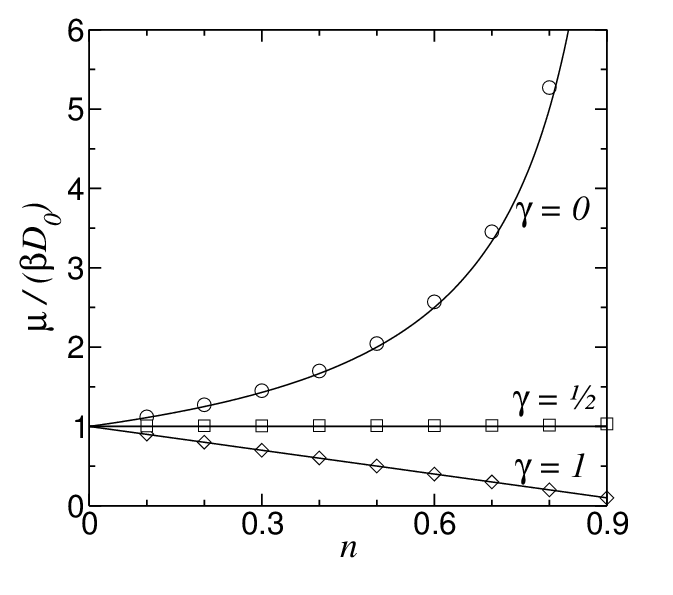}
\caption{Fermion's mobility $\mu$ against concentration $n$ for different values of $\gamma$, obtained from a system with homogeneous concentration, constant force $\beta F a=0.05$ and periodic boundary conditions (see text for more details). The curves correspond to Eq.\ \eqref{mu}. Dots correspond to Monte Carlo simulations with $\gamma=0$ (circles), $\gamma=1/2$ (squares) and $\gamma=1$ (diamonds). Number of samples: 1000; MC steps: between $10^5$ and $10^6$; $N=100$; system size $L=100 a$.}
\label{mufermions}
\end{figure}

Figure \ref{eqfemions} shows the equilibrium solution with a constant force to the right in a closed system, and Monte Carlo simulation results for different values of $\gamma$ (0, 1/2 and 1); as expected, in all cases the same equilibrium solution is obtained.

As mentioned before, the different values of $\gamma$ become relevant in non-equilibrium situations. We analyzed two simple cases. The first one is a constant force ($U=-F\,x$) applied to a system with periodic boundary conditions. The non-equilibrium stationary state is homogeneous with non-zero current. From Eq.\ \eqref{ccurrent} we can obtain the mobility: $\mu = J\,a/(n\,F)$. Figure \ref{mufermions} shows numerical simulation results of the mobility against concentration for different values of $\gamma$. The results coincide with the analytical curve \eqref{mu}. The analytical expressions of $\mu$ and $D$, for fermions and bosons (next section) and for different values of $\gamma$, are shown in table \ref{tabla}.

\begin{figure}
\includegraphics[width = 0.45\textwidth]{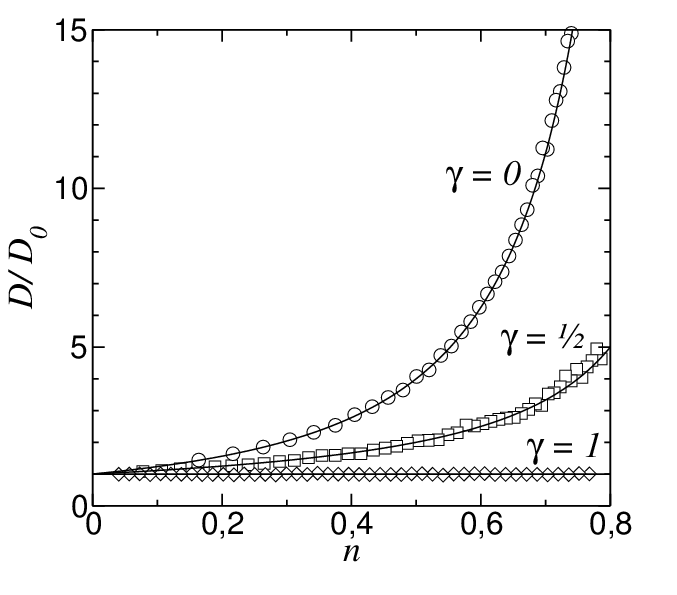}
\caption{Fermion's diffusion coefficient $D$ against concentration $n$ for different values of $\gamma$, obtained from a system with unequal fixed conditions at the ends and zero force, in a non-equilibrium stationary state (constant current $J$, see text for more details). The curves correspond to Eq.\ \eqref{diffcoef}. Dots correspond to Monte Carlo simulations with $\gamma=0$ (circles), $\gamma=1/2$ (squares) and $\gamma=1$ (diamonds). Number of samples: 1000; MC steps: between $10^8$ and $10^9$; $N=500$; system size $L=100 a$; $n(0)=0.8$ and $n(L)=0$.}
\label{differmions}
\end{figure}

The second non-equilibrium situation that we considered is zero force with unequal fixed boundary conditions. The difference in particle concentration at both ends of the system produces a constant current in the stationary state. Numerically, we can obtain the density profile and also its space derivative. From the equation for the current \eqref{ccurrent}, we have that the diffusion coefficient is $D = - J\,a/\frac{\partial n}{\partial x}$. In this way, we can plot the diffusion coefficient against concentration for different values of $\gamma$, as shown in figure \ref{differmions}. The numerical results, again, coincide with the analytical results \eqref{diffcoef} (see table \ref{tabla}).

\begin{table}[ht]
  \caption{\label{tabla}%
Mobility and diffusion coefficient for different values of $\gamma$, as derived from equations \eqref{mu} and \eqref{diffcoef}, for fermions (minus sign) and bosons (plus sign).}
  \begin{ruledtabular}
    \begin{tabular}{lcc}
    \textrm{$\gamma$}&
    \textrm{$\mu/(D_0 \beta)$}&
    \textrm{$D/D_0$}\\
    \colrule
    0 & $1/(1\pm n)$ & $1/(1\pm n)^2$ \\ 
    1/2 & 1 & $1/(1\pm n)$ \\ 
    1 & $1 \pm n$ & 1 \\ 
    \end{tabular}
  \end{ruledtabular}
\end{table}

\begin{figure}
\includegraphics[width = 0.45\textwidth]{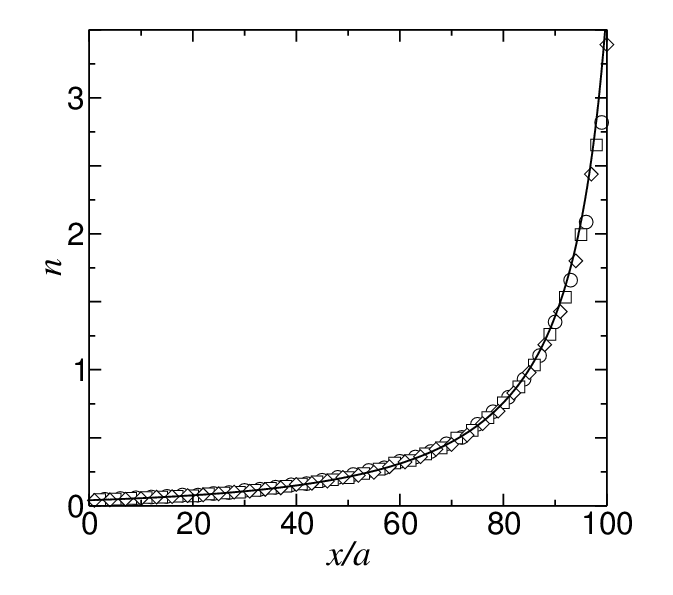}
\caption{Equilibrium concentration $n$ for bosons with energy $U(x) = -F\,x$, $\beta F a=0.03$ and total concentration $0.5$. The curve corresponds to the Bose-Einstein distribution and the dots to Monte Carlo simulations with different values of $\gamma$: $\gamma=0$ (circles), $\gamma=1/2$ (squares) and $\gamma=1$ (diamonds). In order to reduce simulation fluctuations, we define the concentration $n_i$ at site $i$ as $n_i = m_i/N$, where $m_i$ is the number
of particles in this site. Number of samples: 1000; MC steps for each sample: $10^7$; $N=10$; fixed boundary conditions; system size $L=100 a$.}
\label{eqbosons}
\end{figure}

\begin{figure}
\includegraphics[width = 0.45\textwidth]{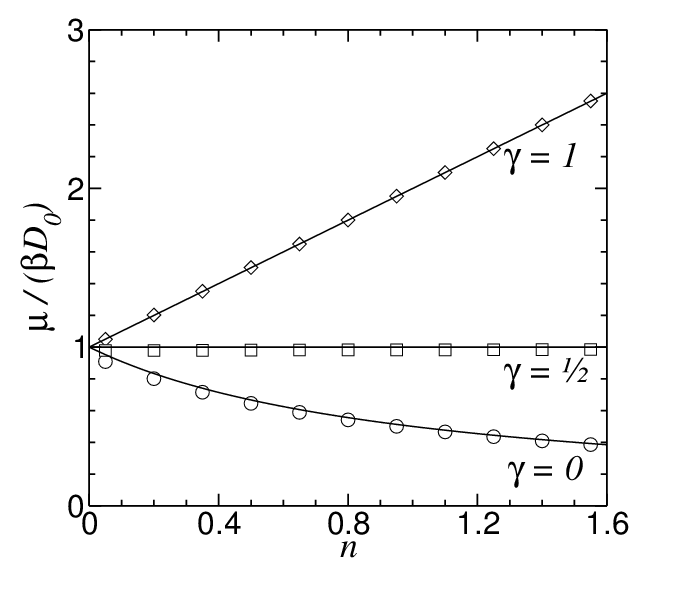}
\caption{Boson's mobility $\mu$ against concentration $n$ for different values of $\gamma$, obtained from a system with homogeneous concentration, constant force $\beta F a=0.05$ and periodic boundary conditions. The curves correspond to Eq.\ \eqref{mu}. Dots correspond to Monte Carlo simulations with $\gamma=0$ (circles), $\gamma=1/2$ (squares) and $\gamma=1$ (diamonds). Number of samples: 5000; MC steps: between $10^5$ and $6\;10^6$; $N=20$; system size $L=100 a$.}
\label{mubosons}
\end{figure}

\section{Bosons}

Boson statistics can be obtained in a classical context using a statistical potential. The behavior of quantum non-interacting particles can be reproduced by classical particles with this effective attractive interaction, that is given by
\begin{equation}
\Phi(n_i) = -\beta^{-1} \ln n_i!
\end{equation}
Its effect is to cancel the Gibbs factor in the partition function \eqref{partfunc}, from which the Bose-Einstein distribution is obtained. 

It is not difficult to obtain the corresponding mean field potential from \eqref{vvsphi}:
\begin{equation}
V(n) = -\beta^{-1}\ln (1+n).
\label{vboson}
\end{equation}
The equilibrium Bose-Einstein distribution is recovered when this result is replaced in \eqref{neq}: $n_\textrm{eq} = 1/(e^{\beta (U-\mu_c)} - 1)$. Figure \ref{eqbosons} shows this equilibrium solution with a constant force to the right. The figure also presents numerical results for different values of $\gamma$, showing that in all cases the same solution is attained.

\begin{figure}
\includegraphics[width = 0.45\textwidth]{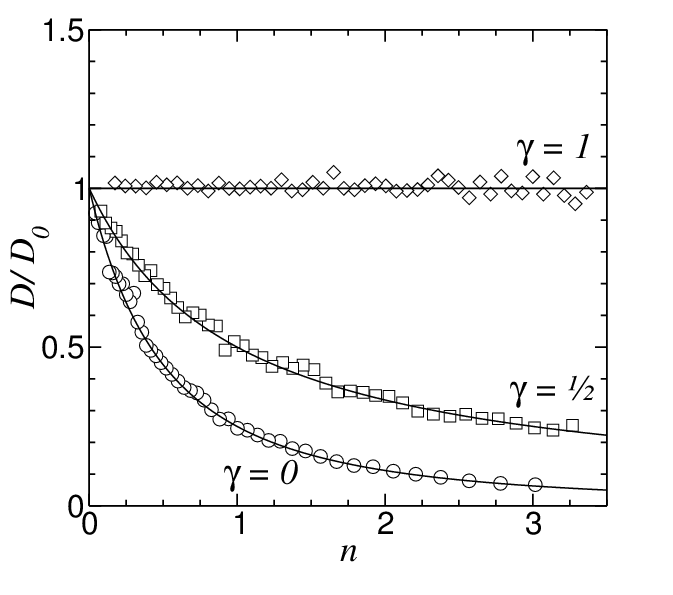}
\caption{Boson's diffusion coefficient $D$ against concentration $n$ for different values of $\gamma$, obtained from a system with unequal fixed conditions at the ends and zero force, in a non-equilibrium stationary state. The curves correspond to Eq.\ \eqref{diffcoef}. Dots correspond to Monte Carlo simulations with $\gamma=0$ (circles), $\gamma=1/2$ (squares) and $\gamma=1$ (diamonds). Number of samples: 5000; MC steps: between $10^7$ and $10^8$; $N=100$; system size $L=100 a$; $n(0)=3.5$ and $n(L)=0$.}
\label{difbosons}
\end{figure}

As in the case of hard core interaction (or fermions), the mean field potential determines the equilibrium solution but not the dynamics. The mobility and diffusion coefficient are not unequivocally determined by $V$, they depend also on $\gamma$. As in the previous section, we analyzed two  non-equilibrium steady state situations from which the mobility and the diffusion coefficient against concentration can be obtained. Numerical results coincide with the corresponding equations \eqref{mu} and \eqref{diffcoef}, as shown in figures \ref{mubosons} and \ref{difbosons}. 

In all cases, for fermions or bosons, in the limit of low concentration, both the mobility and the diffusion coefficient coincide with the corresponding non interacting values: $\beta D_0$ and $D_0$ respectively.

\section{Generalizations}

The generalization of the nonlinear Fokker-Planck equation \eqref{fp} to higher dimensions is straightforward:
\begin{equation}
\frac{\partial n}{\partial t} = \nabla \cdot \left( \mu\,  \nabla U\, n +  D\, \nabla n \right).
\label{fphd}
\end{equation}
The expressions for $\mu$ and $D$, \eqref{mu} and \eqref{diffcoef}, remain unchanged. The relation between $V$ and $\Phi$, \eqref{vvsphi}, also keeps its validity for higher dimensions. Let us note that these equations also hold for a space dependent $\gamma$ parameter.

For the generalization to continuous systems we have to take into account some considerations. The system is divided in cells of size $a$, small enough to be considered point-like and, at the same time, large enough to contain many particles (a standard approach in non-equilibrium statistical mechanics). The cell size is much smaller than a typical concentration wavenumber $\lambda$. Also, the interaction range $r$ should be much smaller than the cell size. So, we have $r \ll a \ll \lambda$. 

Condition $a \ll \lambda$ allows us to take the continuous limit and obtain the Fokker-Planck equation. And condition $r \ll a$ allows us to neglect the interaction between cells, since the interaction energy in the cell's surface is much smaller than in the bulk, and to keep the validity of the relation \eqref{vvsphi} between $V$ and $\Phi$. The consequence is that the Fokker-Planck equation is local.

\section{Conclusions}

We derived a non-linear Fokker-Planck equation for interacting particles. The derivation is based on an energy landscape of wells and barriers on a lattice. The mean field potential and the external potential determine the energy wells and the equilibrium solution. The barrier's heights depend on the $\gamma$ parameter; it determines if the transition probability depends on the mean field potential of the origin site, the target site, or on a mixture of both. 

A relation between the mean field potential and the microscopic interaction energy was deduced. The Jarzynski equality can be used to interpret the mean field potential as the free energy change when the work needed to increase the number of particles by one is applied.

The results are illustrated with the hard core (or fermion) and boson interactions. The corresponding mean field potentials can be combined with different values of $\gamma$. The consequence is that, for the same potential and equilibrium solution, the dependence of the mobility and diffusion coefficient on the concentration can have large variations determined by the value of $\gamma$. In all cases, for small concentration, mobility and diffusion coefficient tend to the corresponding non interacting values.

We considered only different constant values of $\gamma$ but, in general, it could be a function of the position or of the concentration.

\end{document}